\documentclass[useAMS,usenatbib]{mn2e}

\usepackage{graphicx}
\usepackage{graphics}
\usepackage{mathtools}
\usepackage{url}

\newcommand{\dd}{\mathrm{d}}
\newcommand{\AAA}{\AA$\,\,$}
\newcommand{\ith}{$i$th }

\begin{document}

\title[Modelling the C-Si charge exchange]{Carbon Cations and Silicon Atoms in the ISM: modeling  their charge exchange reaction }

\author[M. Satta et al.]{M. Satta$^1$, 
	T. Grassi$^2$, and F. A. Gianturco$^2$\thanks{Corresponding author: e-mail: fa.gianturco@caspur.it Fax +39-06-49913305}\\
$^{1}$CNR-ISMN and University of Rome "Sapienza", P. le A. Moro 5, 00183, Rome, Italy\\
$^{2}$Dept. of Chemistry, University of Rome ``Sapienza'', P.le A. Moro 5, 00161 Rome, Italy\\}

\date{Accepted *****. Received *****; in original form ******}

\pagerange{\pageref{firstpage}--\pageref{lastpage}} \pubyear{2012}

\maketitle
\label{firstpage}

\begin{abstract}
The time-dependent rate coefficients for the charge exchange reaction $\mathrm{C}^+ + \mathrm{Si}\to\mathrm{C} + \mathrm{Si}^+$ for doublet and quartet states have been determined with \emph{ab initio} quantum calculations  coupled with a non-adiabatic transition model based on a simple Landau-Zener picture. This reaction plays a key role in determining the abundances of C, Si, and their ions, in the ISM since these abundances affect the fine structure cooling
and hence the star formation rates. We also provide additional calculations  to evaluate the differences between the gas evolution as obtained by using the empirical rate estimates found in the current literature and the calculations presented in this work which are based on our more realistic evaluation of such rates from \emph{ab initio} transition probabilities. We shall thus show here  that the new rates yield important differences for metal-rich environments where $T<10^4$ K and the
UV flux is very small, while becoming less important at higher T values and higher photon fluxes. 
\end{abstract}

\begin{keywords}
astrochemistry -- ISM: evolution, atomic
\end{keywords}

%--------------------------------------
%--------------------------------------
%--------------------------------------
\section{Introduction}
The chemical transformations occurring in the Interstellar Medium (ISM) and protoplanetary atmospheres  plays a key role in explaining the evolution and history of its objects (e.g. \citealt{GalliPalla98}, \citealt{Nelson1999},  \citealt{MerlinChiosi07}, \citealt{Tesileanu2008}, \citealt{Gnedin2009}, \citealt{Glover2010}, \citealt{Yamasawa2011}, \citealt{Grassi2011}), since they allow  to determine the total amount of their cooling and hence to derive specific information on the star formation of a given ISM region. Chemistry is also important for providing explanations and helping the  comparison between the observations through spectroscopic experiments (e.g. \citealt{Furuya2012, Combes2012}) and their computed simulations. Our chemical data are  also employed to compute the optical depth of the various regions of the  ISM, thereby determining the amount of photons that influence the evolution of a given environment (e.g. \citealt{Latif2011,Petkova2012}).

Unfortunately, including an accurate and complete network of the chemical processes when endeavouring to perform an   ISM environment evolutionary simulation 
has two main drawbacks:  the first one has a computational origin and it comes from the mathematical nature of the problem, which is currently represented 
by a system of first-order differential equations (ODE). Solving coupled  ODEs usually has  a non-negligible computational cost for a series of reasons which are, however, beyond the aims of the present work so that we refer any interested reader to our recent study of this problem(see \citealt{Grassi2012}).
The second problem will be specifically discussed in this paper, of which it constitutes the main task, and is related to the accuracy of the rate coefficients $k(T)$ of the molecular processes and chemical reactions usually employed in the model calculations of object evolutions in the ISM. For a given reaction $\mathrm{A}+\mathrm{B}\to\mathrm{C}+\mathrm{D}$, in fact,  the rate per unit volume $F$ (i.e. number density per unit of time) is determined by the quantity $F=k(T)\,n_\mathrm{A}\,n_\mathrm{B}$, where $n$ is the number density of the species indicated by the subscripts.  Each reaction has therefore its own $k(T)$ that determines the efficiency of the process, and all the rates together affect the evolutionary model specifically employed. It is thus rather straightforward to understand  that to properly model the ISM one needs as many accurate rates as possible and that  the accuracy depends on the method used to estimate the desired $k(T)$.
There are four widely-used methods to obtain a reaction rate: (i) one of the most accurate is to have access to experimental results, but
unfortunately not all the reactions can be obtained trough this method, often because these experiments are either not feasible or not available at a given time. 
(ii) Another way is to use the Langevin model (e.g. see \citealt{SteinfeldBook1989}, Sect.8.3) that represents a crude approximation to any reaction which would involve the interaction of an ion with a neutral species, for which is uses an adiabatic capture model on a single potential energy surface. It is not very accurate and, moreover, it misses the temperature dependence that is often a crucial property of ion-molecule reactions. 
(iii) The third approach is based on employing realistic chemical assumptions to describe reactions for which we have no direct information but are similar to others which we know more directly. An example would be an  isotopic substitution or similar (for example considering that HD+H behaves the same as H$_2$+H). This method can be a necessary approximation but it could obviously lead to large errors. 
(iv) The fourth option is the one we follow  in this paper and consists in employing an  \emph{ab initio} method which allows us to obtain accurate reaction rates via quality-controlled theoretical calculations that could in principle be treated as exact procedures. This approach gives rate coefficients which are usually expected to carry fairly small errors, but it is usually  more likely to be possible  for small interacting systems. The   one we shall be discussing below constitutes a specific example of such a procedure.

The process we  are considering  in this paper plays a key role when we study ISM environments with a rather weak local UV photon flux (e.g. where the gas is optically thick) and therefore the metals of such regions are not fully ionized because of the small number of ionizing photons. 
On the other hand,  at the same time they are not completely neutral since there are other ionizing processes such as collisions and reactions with cosmic rays.
For these reasons when the gas is optically thick we shall have to take into account the  $k_\mathrm{CE}(T)\,n_\mathrm{C^+}\,n_\mathrm{Si}>0$ where $k_\mathrm{CE}(T)$ is the  reaction rate coefficient for the process
\begin{equation}
	\mathrm{C}^+ + \mathrm{Si} \xrightarrow{k_\mathrm{CE}} \mathrm{C} + \mathrm{Si}^+\,.
\end{equation}
The importance of determining the abundances of the different metals (and ions) is related to the fact that at a temperature lower 
than $10^4$ K the cooling is dominated by the fine-structure cooling of the metals \citep{Maio07, Grassi2011}. 
The ions of a given metal have a different cooling efficiency  from that of its corresponding neutral partner \citep{Santoro2006},
so that to employ in any model the correct value of $k_\mathrm{CE}(T)$ will affect the total cooling of the gas which is produced by that very  model.
In Sect.\ref{sect:calculations} we shall discuss the computational methods which we employ to determine the reaction rates of the charge exchange processes between C$^+$ and Si and its counterpart couple Si$^+$ and C. We will also provide there an analytical  fit for their relevant  $k_\mathrm{CE}(T)$. In Sect.\ref{sect:resudiscu} we further analyse the behaviour of the non-adiabatic crossings between the potential energy curves for both quartet and doublet states of the combined system, we additionally present our findings for  the cross-sections and rate constants of the charge exchange process over a range of temperatures and we finally  compute their effects within an evolutionary model and discuss  the differences which we find between the evolution of the ISM  including the only existing empirical estimate for the relevant  reaction rate \citep{LeTeuff2000} and that evolutionary behaviour obtained by employing the more realistic and accurate rates which have been produced  in the present work. 

%--------------------------------------
%--------------------------------------
%--------------------------------------
\section{Theoretical method}\label{sect:calculations}

The diatomic molecular cation SiC$^+$ has been the subject of \emph{ab initio}
 studies since the work of \citet{Bruna1981}, where the dissociation channel leading to C$(^3$P$_g)+$Si$^+(^2$P$_u)$ was investigated through the potential energy curves relative to the electronic quartet and doublet states within the multireference correlation method given by the  MRD-CI approach. More recently \citet{Pramanik2008} have employed multireference singles and doubles configuration interaction to study the electronic structures and spectroscopic properties of the SiC$^+$ cation. They have calculated all the potential energy curves leading to C$(^3$P$_g)+$Si$^+(^2$P$_u)$ and two out of nine of the doublet potential energy curves which dissociate into the excited electronic state C$(^1$D$_g)+$Si$^+(^2$P$_u)$. We have used their data relative to the quartet states as a part of the total potential energy curves which we shall employ below to calculate the charge transfer cross-sections and rate constants. 
The doublet potential energy curves for the Si$^+/$C states have been computed within the Complete Active Space SCF (CASSCF) methodology set up by \citet{werner85}.  The active space was given by nine molecular orbitals and seven electrons were kept active during the energy calculations. The state-averaged procedure have been used to overcame the root-flipping problems: in particular the n-th root is calculated with a unitary weigth, and 0.01 the weigths of all the lower energy states. The cc-pVTZ correlation-consistent polarized basis set developed by \citet{Dunning1989} was employed for these excited state calculations and the C$_{2v}$ symmetry was imposed to the molecular orbitals.  Due to convergence problems, the CASSCF procedure, failed in producing five out of the  nine doublet potential energy curves associated with C$(^1$D$_g)+$Si$^+(^2$P$_u)$, and the quartet and doublet ground states associated 
with Si$(^3$P$_g)+$C$^+(^2$P$_u)$. 
A different approach was therefore adopted for the calculations of the quartet and doublet potential energy curves whose 
asymptote is Si$(^3$P$_g)+$C$^+(^2$P$_u)$.
These last two curves have been calculated within  the framework of the constrained Density Functional Theory (DFT) formalism of \citet{Wu2005}, using the B3LYP density functional \citep{Becke1993} and building the starting wavefunction with the positive charge confined to be on the carbon atom. The potential energy curves calculated with these two different methods have been asymptotically shifted in order to reproduce the difference in the ionization potential of the C and Si atoms, while the differences in the interaction regions were obtained from our calculations.  The CASSCF calculations were performed using the MOLPRO \citep{MOLPRO_brief} package, while  the NWCHEM program \citep{Valiev2010} was used to compute the DFT potential energy curves. The range of the carbon-silicon distances spanned was chosen to be between 1.25 and 6.25 \AAA with individual  steps of 0.1 \AA.

The probability for the occurrence of  the charge exchange process \citep{Johnson1982} after  every \ith crossing between  the potential energy curves of different atomic charge localization has been calculated within  the Landau-Zener-Stueckelberg (LZS) approximation \citep{Landau1932, Zener1932, Stueckelberg1932}:
\begin{equation}
 P^i_{0f}=\exp\left(\frac{-\tau^i_{x}}{\tau^i_{0f}}\right)\,,
\end{equation}
where $\tau^i_{x}=\frac{\delta R^i}{\left|v^i(R_{x_i})\right|}$ and $\tau^i_{0f}=\left[\pi\left|V^i_{0f}\right|_{R_{x_i}}\right]^{-1}$. The term $R_{x_i}$ is the interatomic distance at which the \ith crossing occurs, $v^i(R_{x_i})$ is the relative velocity between the two atoms at the \ith crossing point. 
We define $\delta R^i=2\left|\frac{^iV_{0f}}{^iV_{ff}^{'}-^iV_{00}^{'}}\right|_{R_{x_i}}$ which depends on the derivatives of the two crossing curves $^iV_{ff}^{'}$ and $^iV_{00}^{'}$ at the  \ith crossing point. Moreover, $^iV_{0f}$ is the  \ith non-resonant exchange interaction and depends on the effective ionization energies, and on the  \ith crossing point coordinate:
\begin{equation}
  ^iV_{0f}=V_{0}\beta R_{x_i}e^{-\beta R_{x_i}}\,,
\end{equation}
with $V_{0}=\sqrt{I_\mathrm{Si}}\sqrt{I_\mathrm{C}}/0.86$, $\beta=0.86\left(\sqrt{I_\mathrm{Si}}+\sqrt{I_\mathrm{C}}\right)/\sqrt{2}$. The effective ionization energies have been calculated as a function of the effective nuclear charge $Z_\mathrm{eff}$ as $I=0.5Z_\mathrm{eff}^{2}$. We have considered only the core electrons as effectively shielding the nucleus, and have been using the data of \citet{Clementi1963} to obtain a final  effective nuclear charge of 3.2 and 3.3 for carbon and silicon respectively.
The \ith charge exchange probability is then employed in turn  to calculate the corresponding \ith cross-section:
\begin{equation}\label{eqn:sigma}
 \sigma^i(v)=\pi R_{x_i}^{2}P^i_{0f}(1-P^i_{0f})\,.
\end{equation}
The final cross-sections are then  summed over all the crossing processes and further integrated over their relevant velocities  to obtain the rate constants 
\begin{equation}\label{eqn:rate}
k_{CE}(T)=4\pi\left(\frac{\mu}{2\pi KT}\right)^{\frac{3}{2}}\int\sum_{i}{\sigma^i(v)}v^{3}\, e^{-\frac{\mu v^{2}}{4KT}}\dd v\,.
\end{equation}
The latter quantities will then be used within the evolutionary modeling described in the next Section.

The above averaging over velocities also ensures that possible shifts on the crossing positions, when using different ab initio curves, are spread over the required velocity values and therefore made less sensitive in getting the final cross sections.

\section{Results and Discussion}\label{sect:resudiscu}
\subsection{Potential Energy Curves and non-adiabatic curve crossings}\label{subsect:quantum}
\begin{figure}
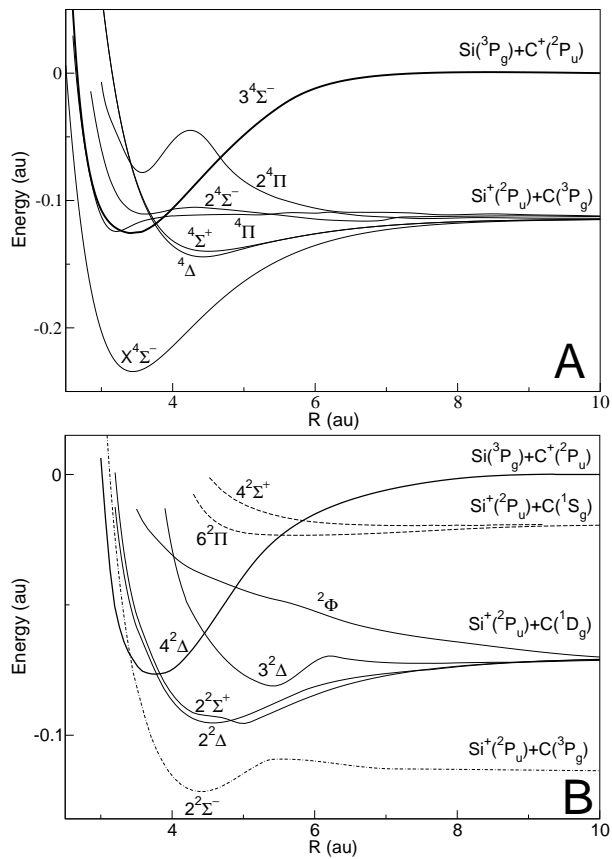

\includegraphics[width=.45\textwidth]{figs/quartet.au.eps} %\plottwo
\includegraphics[width=.45\textwidth]{figs/doublet.nw.eps} %\plottwo
\caption{Potential energy curves for doublet (bottom) and quartet (top) states of carbon-silicon charge exchange. The C/Si$^+$ and C$^+$/Si curves have been calculated by CASSCF and constrained-DFT method respectively. See text for details on their calculations.\label{fig:DQ}}
\end{figure}

The process of charge transfer from the cation C$^+$ to the neutral partner Si involves all the potential energy curves whose asymptotic energy levels are included between the two electronic states with opposite atomic charge localization:  C$(^3$P$_g)+$Si$^+(^2$P$_u)$ and  
Si$(^3$P$_g)+$C$^+(^2$P$_u)$. In the case of the quartet states all the energy levels asymptotically converge to either  one of the two charge transfer boundary levels shown above (see Fig.\ref{fig:DQ} top). The doublet states, on the contrary, present two electronic excited levels embedded within the two charge exchange boundary levels, namely the C$(^1$D$_g)+$Si$^+(^2$P$_u)$ and  C$(^1$S$)+$Si$^+(^2$P$_u)$ (see Fig.\ref{fig:DQ} bottom).
\indent In order to have a more complete picture of the potential energy curves which control the charge transfer process we have calculated, besides the doublet and quartet electronic ground states associated with Si$(^3$P$_g)+$C$^+(^2$P$_u)$, also the $2^2\Delta$, the $2^2\Sigma^+$, the $3^2\Delta$ and the $^2\Phi$ dissociating into C$(^1$D$_g)+$Si$^+(^2$P$_u)$ and the $6^2\Pi$ and $4^2\Sigma^+$ associated with C$(^1$S$)+$Si$^+(^2$P$_u)$. We were not able, due to instabilities in the convergence process, to compute the other four doublet levels whose asymptote is C$(^1$D$_g)+$Si$^+(^2$P$_u)$. All the present calculated  data turn out to be in good agreement with the earlier results of \citet{Pramanik2008}.

Among  the quartet states there are seven single crossings between the potential energy curve for which the positive charge is localized on the carbon atom (black thick line of Fig.\ref{fig:DQ} top), and the other five potential energy curves, while  the $X^4\Sigma^-$ does not participate to the charge exchange process. The crossings are seven because the $^4\Pi$ crosses three times the C$^+$/Si curve, in the repulsive region at $R_x=2.81$ au and in the diatomic ion bound region at $R_x=3.27$ au and $R_x=3.88$ au. The $^4\Delta$ and the $^4\Sigma^+$ are quasi-degenerate states, and cross the C$^+$/Si curve in their repulsive parts, the other crossing occurs at the larger internuclear distance of 4.67 au.  

The doublet states crossings involve only the higher energy curve ($2^2\Sigma^-$) of the six curves which dissociate into the ground electronic levels of the cation Si$^+(^2$P$_u$) and neutral carbon atom ($^3$P$_g$). The crossing points, which occur at larger internuclear distances, involve the C$^+$/Si curve in its attractive part, and the four curves, in their repulsive regions, dissociating into C$(^1$D$_g)+$Si$^+(^2$P$_u)$. The two higher energy crossing points are characterized by internuclear distances larger than 5 au, and their energies are about 0.5 eV below the entrance channels C$^+$/Si.

\begin{figure}
\includegraphics[width=0.45\textwidth]{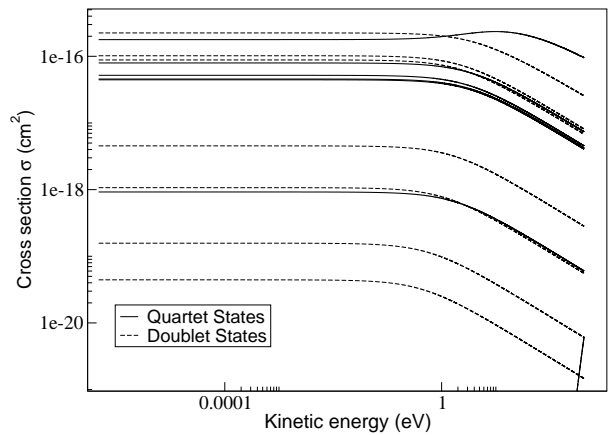}
\caption{Cross-sections for the charge exchange reaction in the doublet and quartet states of SiC$^+$. See text for details.\label{fig:sigma}}
\end{figure}

The structural and energetic parameters (interatomic distances, energy locations and their derivatives) have been employed to calculate the charge exchange cross-sections $\sigma^i$ (see. Eqn. \ref{eqn:sigma}) for all the crossings involving  both quartet and doublet states. In Fig.\ref{fig:sigma} we report the cross-sections as a function of the relative kinetic energy of the Si and C atoms. The behaviour of these $\sigma$ is almost constant below 1 eV while at higher kinetic energies, from 1 eV to 400 eV, there is a marked decrease in the cross-sections with an exponential decay factor which is almost the same for all the $\sigma$. An exception is the behaviour of the cross-section associated with the quartet state $^4\Pi$ at $R_x=2.81$ au (bottom line of Fig.\ref{fig:sigma}). In this case the low values of the $\sigma$ are due to the short range at which the charge exchange process can take place. A fact connected with the short range repulsion forces which therefore allow the interaction of the electrons with their nuclei to get stronger, thereby increasing localization and preventing electron charge  to easily flow from one atom to the other. A different instance is provided by the crossing of 
$\,^4\Pi$ at 3.27 au which produces a final cross-section that increases up to 10 eV and then decreases exponentially,  similarly to the other $\sigma$ originating from crossings between the other states. This different behaviour can be explained in terms of the relevant potential energy curves proceeding very smoothly, and nearly parallel to each other, into the crossing region, a feature which shifts the maximum of the $\sigma$ to the higher kinetic energy with respect to the other charge exchange cross-sections between those potential energy curves that reach more abruptly their crossing points.
 
\begin{figure}
\includegraphics[width=.45\textwidth]{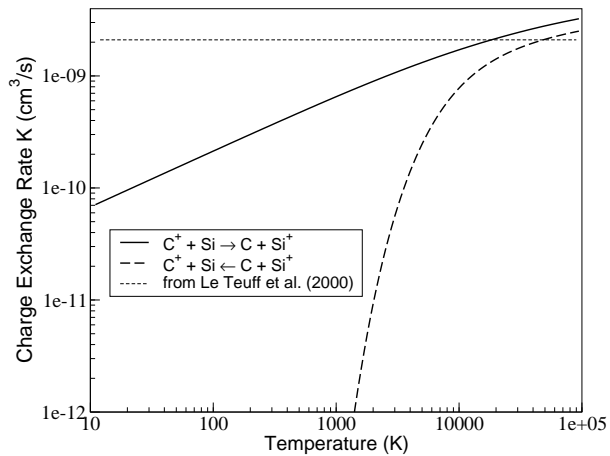}
\caption{The rate coefficient calculated in this Paper (solid line) compared to the one in \citet{LeTeuff2000}
(thin dashed line). We also include the inverse process (thick dashed line).\label{fig:rate}}
\end{figure}

The presently computed final cross-sections have been employed to calculate the corresponding rate constants for the charge exchange process of interest,  following Eqn.(\ref{eqn:rate}), as shown in Fig.\ref{fig:rate}. We also reported there  the rate constants of the inverse charge exchange, in which the electron density flows from the carbon atom to the silicon cation. In Fig.\ref{fig:rate} we also report for comparison the rate constant which is usually employed in the ISM modelling, taken from UMIST database\footnote{\url{http://www.udfa.net/}} \citep{LeTeuff2000}.   We clearly see there that the newly computed rate constants for the direct process (C$^++$Si$\to$C$+$Si$^+$) are several orders of magnitude larger than those for the inverse process (C$^++$Si$\to$C$+$Si$^+$), and only at temperatures becoming higher than 5000 K the two rate constants begin to converge to very similar values. On the other hand, the earlier estimate for this  rate constant of  \citet{LeTeuff2000} does not depend on the temperature and therefore it is clearly higher than the rate constant calculated in this work at temperatures which are larger  than about $1.1\times10^4$ K, while at the lower temperatures shown in the figure the present  rate constants are more than an order of magnitude  smaller than those of the earlier empirical estimates \citep{LeTeuff2000}.

We have also generated a fitting function that analytically reproduce  the results of our calculation as a function of T
\begin{equation}
	k_\mathrm{CE}=\left[1.8679\times10^8 + 5.091022\times10^{10}\,T^{-0.526561}\right]^{-1}
\end{equation}
in the range $T\in[10,10^5]$ K.

%--------------------------------------
%--------------------------------------
%--------------------------------------
\subsection{Modeling the evolution of the ISM: a comparison}\label{subsect:models}

\begin{figure*}
\includegraphics[width=.4\textwidth]{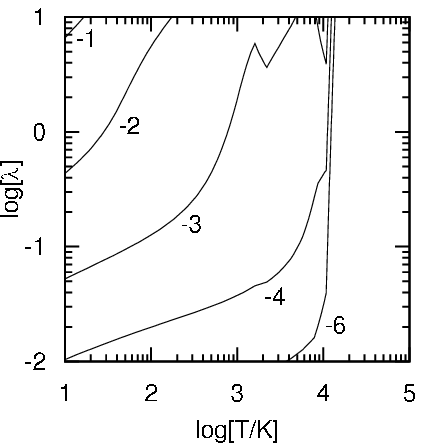} %\plottwo
\includegraphics[width=.4\textwidth]{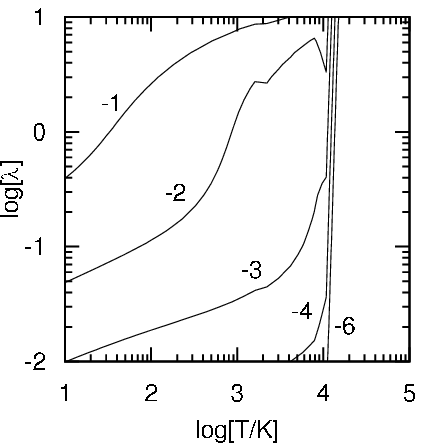} %\plottwo
\caption{Contour plot of logarithm of the normalized fraction defined in Eqn.(\ref{eqn:def_frac}) for \emph{low} (left) and \emph{high} (right) models. 
Contour values are indicated by the labels. 
See text for details.
\label{fig:low_high_Cp}}
\end{figure*}

Our newly obtained reaction rates have been included in a one-zone ISM model to test their direct influence over the corresponding evolution of the gas and relation with the influence already surmised in the same modelling but using the earlier estimates of the same charge exchange  rates \citep{LeTeuff2000}. 
We define $k_\mathrm{old}$ as the rate coefficient indicated in \citet{LeTeuff2000}, while $k_\mathrm{new}$ as the one provided by our
calculations. In the model presented in this Section we follow the time-evolution of the abundances of the chemical species given a set of initial 
abundances, and a grid of temperature ($T$) and metal fraction ($\lambda$) as described below. The chemical network is based on the more general one of \citet{GloverJappsen2007}, and we solve the system of differential equations associated to this network with a standard DLSODES solver \citep{Hindmarsh83}. 
We let the system evolve for $10^4$ yr keeping the temperature and the total density both constant.

We found that the new rates change the behaviour of the ISM gas only 
in the absence  of a UV photon source, since the presence of such an intense source would massively increase the number of both Si and C ions produced by direct photoionization, thereby   reducing the efficiency of the charge exchange due to the rapid disappearance  of neutral partners in that reaction (i.e. Si).
By keeping the latter assumption, we have therefore tested our model when selecting a thick gas environment over a grid of $T$ and $\lambda$ values, in particular by changing the gas temperature within the range $T=[10,10^5]$ K, with different initial number densities of the  metals present in the gas according to the following expression: 
$n_\mathrm{X}=\lambda\,n_\mathrm{Htot}\,\hat n_{\mathrm{X}}$, where $\hat n_{\mathrm{X}}$ is the default abundance of the metal X, 
$n_\mathrm{Htot}=n_\mathrm{H^+} + n_\mathrm{H} + 2 n_\mathrm{H_2}$ is the total abundance of hydrogen, while $\lambda = [0.1,10]$. 
Following \cite{Wakelam2008} we have also assumed  $\hat n_\mathrm{He}=9\times10^{-2}$, $\hat n_\mathrm{O}=2.56\times10^{-4}$, 
$\hat n_\mathrm{C^+}=1.2\times10^{-4}$, where all the abundances are in cm$^{-3}$. For silicon atoms  we have selected  two different cases:
 $\hat n_\mathrm{Si^+}=1.7\times10^{-6}$ cm$^{-3}$ and $\hat n_\mathrm{Si^+}=1.7\times10^{-5}$ cm$^{-3}$ labelled \emph{low} and \emph{high}
model respectively. We have also chosen $n_\mathrm{H}=10^3$ cm$^{-3}$, and $n_\mathrm{e^-}=\sum_{i\in\mathrm{ions}}n_i$ only at the beginning of the simulations (i.e. the initial condition for free electrons). 
Finally, we have set the cosmic rays ionization rate to $\zeta_\mathrm{CR}=1.3\times10^{-17}$ s$^{-1}$.

To determine the differences produced within the model when choosing either  the original rates ($k_\mathrm{old}$) or  the new ones ($k_\mathrm{new}$)
we let the system evolve for $t_\mathrm{end}=10^4$ yr for both cases changing $\lambda$ and $T$ within the preselected range indicated before,
in order to fill the $T-\lambda$ grid of models.
The results are plotted in Fig.\ref{fig:low_high_Cp} for both \emph{low} and \emph{high}
models as the logarithm of the normalized percent difference between 
the two C$^+$ abundances at $t_\mathrm{end}$, namely 
\begin{equation}\label{eqn:def_frac}
	\delta_\mathrm{C^+}=\left.\log\left(\frac{\left|n_\mathrm{C^+}^\mathrm{new}-n_\mathrm{C^+}^\mathrm{old}\right|}{n_\mathrm{C^+}^\mathrm{old}}\right)\right|_{t=t_\mathrm{end}}\,.
\end{equation}

The contour lines in both panel are labelled according to $\delta_\mathrm{C^+}$ (e.g. $\delta_\mathrm{C^+}=-1$ represents a difference of the 
$10\%$ between the value evolved with $k_\mathrm{old}$ and the same for $k_\mathrm{new}$).
More in detail, the two plots represent in the $T-\lambda$ space the difference between the evolved $n_\mathrm{C^+}$ at $t=t_\mathrm{end}$ 
for a grid of models that include $k_\mathrm{old}$ and $k_\mathrm{new}$, and higher contour values represent
higher differences, i.e. the regions of the $T-\lambda$ space where the new rate of carbon-silicon charge exchange 
calculated in this work has the larger effect.

As expected the differences between the models essentially vanish when $T\approx 10^4$ K since at this temperature we have
 $k_\mathrm{old}\approx k_\mathrm{new}$; for the same reason $\delta_\mathrm{C^+}$ increases at the lower temperatures where 
the two rates attain  their largest differences. Furthermore, $\delta_\mathrm{C^+}$ 
is larger for the  larger values of $\lambda$, and also when the abundance of Si is larger,  because the increase of the colliding frequencies between 
partners increases the probability of the  carbon ion to react with the Si partner.

%--------------------------------------
%--------------------------------------
%--------------------------------------
\section{Conclusions}\label{sect:conclusion}
The work which we have discussed in the previous sections is directed at the determination from first principles quantum methods of the dynamics of charge transfer collisions between two atomic partners of the ISM composition, silicon and carbon. In both cases we considered either the C or the Si atom to be the cationic species involved in the collisional process. The quantum mechanical calculations have revealed that, as expected, several ionic potential energy curves (PECs) are involved within the physically interesting range of energies and that several nonadiabatic curve crossing features are taking place, thereby guiding the charge transfer processes in both directions, albeit with different probabilities.
The accuracy of the calculated PECs has been tested in comparison with earlier calculations and found to correctly describe the process of interest. Furthermore, we have modelled the nonadiabatic curve crossing dynamics occurring by collisional energy transfer by using the Landau-Zener-Stueckelberg formulae discussed in one of the preceding sections of this paper. These probabilities can further be used to generate Charge Exchange (CE) cross-sections which, in turn, yield the CE rate constants for the overall process and over a broad range of temperatures.
The only existing estimate of the CE process has briefly presented in  \citet{LeTeuff2000}. It was essentially a Langevin-type of rate calculations where the atomic polarizability that drives the capture rate in the Langevin model (a single potential model) was  replaced by a mass-weighted average of the two polarizabilities of the neutral partners in the entrance and exit channels of the CE process. The main feature of that simple model was to yield rates independent of the temperature of the ISM gas. On the contrary, the present calculations have shown that the CE dynamics is dependent on the partners' velocities at the crossings and therefore generate by necessity a set of temperature dependent rates.
We have further employed the new rates in a numerical modelling of the gas evolution, as also described in detail in the previous section of this work. Our results indicate that in situations of low-metallicity and low photon flux into the gas, the new rates suggest a markedly different behaviour with respect to the previous analysis. On the other hand, when the photon flux increases then the overall photoionization channels dominate the evolution and make the collisional CE path essentially too inefficient.
The present study has therefore provided a clear and specific example of the importance in obtaining realistic descriptions of the chemical processes appearing in evolutionary models so that their final behaviour remains based on more realistic physical pictures.

%--------------------------------------
%--------------------------------------
%--------------------------------------

\section*{Acknowledgements}
The computational support from the CASPUR Consortium is gratefully acknowledged, as well as the financial
support from the PRIN 2009 research network. One of us (T.G.) thanks the CINECA Consortium
for the awarding of a postdoctoral grant during which this work was carried out.

\bibliographystyle{apj}      
\bibliography{mybib} 

\bsp

\label{lastpage}
\end{document}